\documentclass[12pt,fleqn]{article}

\usepackage{graphicx}
\usepackage{pdftricks}

\begin{document}

\title{\Large Thermodynamic curvature for a two-parameter spin model with frustration}

\author{George Ruppeiner\footnote{New College of Florida, Sarasota, Florida, 34243, USA (ruppeiner@ncf.edu)} and Stefano Bellucci\footnote{INFN-Laboratori Nazionali di Frascati, Via E. Fermi 40, 00044 Frascati, Italy (bellucci@lnf.infn.it)}}

\maketitle

\begin{abstract}

Microscopic models of realistic thermodynamic systems usually involve a number of parameters, not all of equal macroscopic relevance. We examine a decorated $(1+3)$ Ising spin chain containing two microscopic parameters: a ``stiff'' $K$ mediating the long-range interactions, and a ``sloppy'' $J$ operating within local spin groups. $K$ dominates the macroscopic behavior, and varying $J$ has weak effect except in regions where $J$ brings about transitions between phases through its conditioning of the local spin groups with which $K$ interacts. We calculate the heat capacity $C_H$, the magnetic susceptibility $\chi_T$, and the thermodynamic curvature $R$. For large $|J/K|$, we identify four magnetic phases: ferromagnetic, antiferromagnetic, and two ferrimagnetic ones, according to the signs of $K$ and $J$. We argue that for characterizing these phases, the strongest picture is offered by the thermodynamic geometric invariant $R$, proportional to the correlation length $\xi$. This picture has correspondences to other cases, such as fluids.

\end{abstract}

\noindent
{\bf Suggested PACS Numbers}: 05.70.-a, 05.40.-a, 64.60.Bd, 75.10.Jm
\\

\par
In microscopic models, the parameters setting the strength of the interactions among the model elements are not usually all equal in importance for determining the overall macroscopic character of the system. Some of these parameters have only a weak influence over the macroscopic properties. The sorting of parameters according to whether they are macroscopically important/unimportant, or ``stiff''/``sloppy'', has recently seen systematic examination in a number of contexts with methods based on the Fisher Information Matrix (FIM) corresponding to the microscopic parameters \cite{Waterfall2006, Machta2013}. The analysis is based on sorting the eigenvalues of the FIM according to their values.

\par
In this paper we propose an extension of these ideas into the thermodynamic realm with a somewhat different FIM, one based on thermodynamic parameters, and resulting from thermodynamic fluctuation theory \cite{Ruppeiner1995, Brody2009}. However, our basic agenda of sorting model parameters according to their effect on the macroscopic behavior is the same in spirit as that of Sethna, {\it et al.} \cite{Waterfall2006, Machta2013}. Our analysis focuses in particular on the invariant thermodynamic Ricci curvature scalar $R$ of the thermodynamic FIM. $R$ reveals information about the character of mesoscopic fluctuating structures. Our viewpoint is that such structures play a significant role in mediating the transition from microscopic to macroscopic, which can be difficult to address with the methods of statistical mechanics \cite{Ruppeiner1995}.

\par
Thermodynamic curvature $R$ is an element of thermodynamic metric geometry. A pioneering paper was authored by Weinhold \cite{Weinhold1975} who introduced a thermodynamic energy inner product. This led to the work of Ruppeiner \cite{Ruppeiner1979} who wrote a Riemannian thermodynamic entropy metric to represent thermodynamic fluctuation theory, and was the first to systematically calculate $R$. A parallel effort was authored by Andresen, Salamon, and Berry \cite{Andresen1984} who began the systematic application of the thermodynamic entropy metric to characterize finite-time thermodynamic processes. $R$ has been worked out in a number of discrete systems \cite{Ruppeiner1981, Janyszek1989, Dolan1998, Janke2002, Dolan2002, Johnston2003, Brody2003, Mirza2013}.

\par
We illustrate our ideas in this paper with a decorated $(1+3)$ Ising spin chain containing two microscopic parameters: a stiff parameter $K$ mediating the long-range interactions, and a sloppy parameter $J$ operating within local spin groups. We show that $K$ dominates the macroscopic behavior, except in cases where varying $J$ brings about transitions between phases through its conditioning of the local spin groups with which $K$ interacts. In addition to $R$, we calculate the heat capacity $C_H$, and the magnetic susceptibility $\chi_T$. We show that $C_H$ is not very effective at displaying the order characterizing the various magnetic phases. $\chi_T$ does a better job, but we argue that $R$ offers the cleanest picture of the magnetic order resulting from $K$. This is the first evaluation of $R$ in a spin model with two coupling parameters.

\par
A strong property of $R$ is that, at zero magnetic field, $R$ is proportional to the correlation length $\xi$ in both the ferromagnetic and the ferrimagnetic phases. Although the model employed here is too simple to fully bring out what Sethna, {\it et al.} \cite{Waterfall2006, Machta2013} have in mind (here the spin groups merely tend to lock into place with each other, instead of having the effects of their local fluctuations averaged out at the mesoscopic level), our use of the terminology stiff/sloppy seems nevertheless appropriate, and sets an agenda for future exploration.

\par
In the theory of critical phenomena, the terms ``relevant" and ``irrelevant" are used for variables which either affect or do not affect universal critical properties \cite{Kadanoff2000}. Our toy model has critical points (at $T = 0$), so we could certainly pitch our discussion in terms of critical phenomena. However, we present our ideas in a broader context, and we get strong results even well beyond what might be termed the critical point regime.

\par
\begin{figure}
\centering
\includegraphics[width=7cm,keepaspectratio=true]{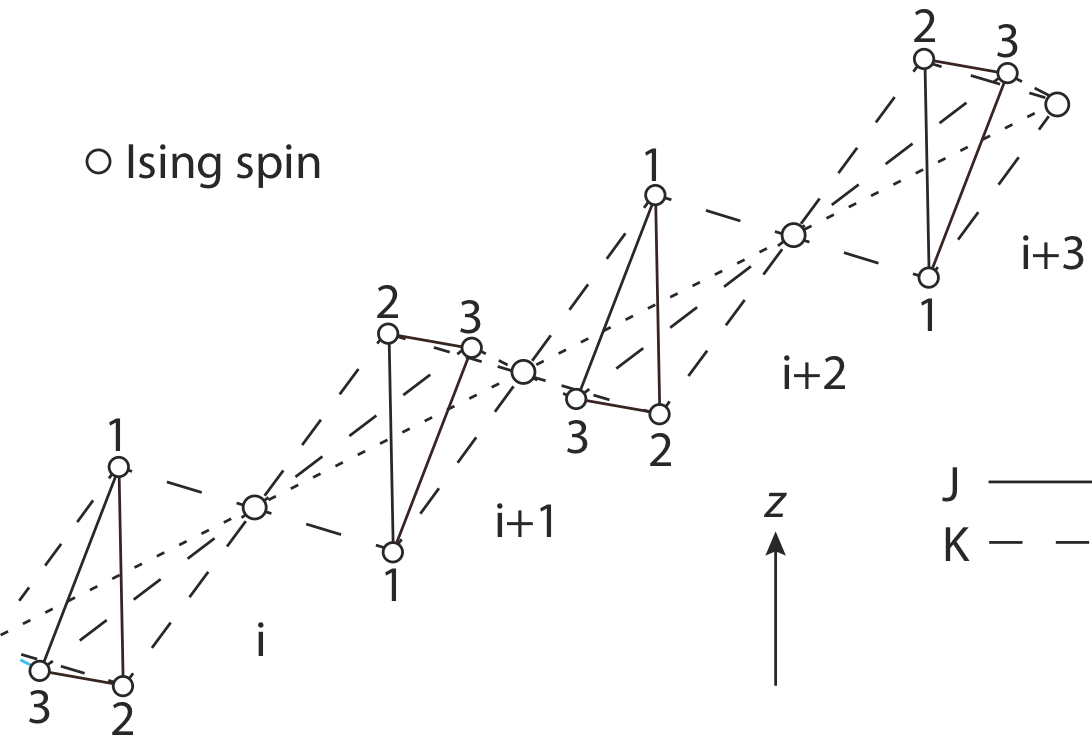}
\caption{The decorated (1+3) Ising chain. The $N$ lattice basis elements, each consisting of a single Ising spin and a three Ising spin plaquette, are enumerated by an index $i\in(1,\dots,N)$, with periodic boundary conditions $(N+1)\leftrightarrow 1$. The plaquette spins are enumerated by an index $\alpha\in (1,2,3)$. Spins within a plaquette interact with each other via a parameter $J$, and with the single neighboring Ising spins via a parameter $K$.}
\label{fig:1}
\end{figure}

Figure \ref{fig:1} shows our spin model, which contains instances of ferromagnetism, antiferromagnetism, and ferrimagnetism. The model consists of $N$ single Ising spins $\sigma_i = \pm 1$, alternating with $N$ triangular Ising spin plaquettes $\sigma_{i\alpha} = \pm 1$. Two such interlaced sublattices offer the possibility of noncanceling magnetic moments, characteristic of ferrimagnetic states \cite{Lieb1962}.

\par
The Hamiltonian $\mathcal{H}$ is a sum over block Hamiltonians $\mathcal{H}_i$,

\begin{equation}\displaystyle\mathcal{H} = \sum\limits_{i=1}^N \displaystyle\mathcal{H}_i,\label{10}\end{equation}

\noindent where

\begin{equation} \begin{array}{lr} \displaystyle\mathcal{H}_i = -\frac{1}{2}H\sigma_i+
 J\left[\sigma_{i1}\sigma_{i2} + \sigma_{i1}\sigma_{i3} +\sigma_{i2}\sigma_{i3} \right]\,+
\\
\displaystyle \,\,\,\,\, K\left(\sigma_i+\sigma_{i+1}\right)\left(\sigma_{i1} +\sigma_{i2}+\sigma_{i3}\right)-H\left(\sigma_{i1} + \sigma_{i2} + \sigma_{i3}\right)- \frac{1}{2}H\sigma_{i+1}, \end{array} \label{15}\end{equation}

\noindent with coupling parameters $(J,K)$, and magnetic field $H$ parallel to the $z$ axis. This block Hamiltonian is that of the solved quantum Ising-Heisenberg chain ($N\rightarrow\infty$) with isotropy parameter $\Delta$ set to zero \cite{Antonosyan2009}. The solution yields the transfer matrix \textbf{T} = $\{\{T_{11}, T_{12}\},\{T_{21} , T_{22}\}\}$ with:

\begin{equation} T_{11} = 2 e^{-h-3 \beta J} \cosh (h+2 \beta K) \left[2 \cosh (2h+4 \beta K)+3 e^{4 \beta J}-1\right],\label{20} \end{equation}

\begin{equation} T_{12} =T_{21} = 6 e^{\beta J} \cosh (h)+2 e^{-3 \beta J} \cosh (3h),\label{30}\end{equation}

\noindent and

\begin{equation} T_{22} = 2 e^{h-3 \beta J} \cosh (h-2 \beta K) \left[2 \cosh (2h-4 \beta K)+3 e^{4 \beta J}-1\right].\label{40}\end{equation}

\noindent Here, $\{\beta,h\} = \{1/T,-H/T\}$, with $T$ the temperature. Boltzmann's constant $k_B = 1$. \textbf{T} has two eigenvalues $\lambda_+$ and $\lambda_-$, ordered as $\lambda_+ > \lambda_-$. The thermodynamic potential per lattice constant (a lattice constant is the distance between spins $\sigma_i$ and $\sigma_{i+1}$) is

\begin{equation}\phi(\beta,h) = \mbox{ln}\,\lambda_+.\label{50}\end{equation}

\noindent $\xi$, in units of lattice constants, for a decorated Ising chain is \cite{Bellucci2013}

\begin{equation}\xi^{-1} = \mbox{ln}\left(\frac{\lambda_+}{\lambda_-}\right). \label{55}\end{equation}

\noindent $\xi$ is nonthermodynamic since it may not be calculated from $\phi(\beta,h)$.

\par
A nice reference model for our discussion consists of a chain of single Ising spins alternating with ``superspins'' $S_i = \pm p$, where $p$ is a positive integer, in place of the triangular spin plaquettes. This model has block Hamiltonian

\begin{equation}\displaystyle\mathcal{H_{S}}_i = -\frac{1}{2}H\sigma_i+K_S(\sigma_i+\sigma_{i+1})S_i-H S_i-\frac{1}{2}H\sigma_{i+1}, \label{60}\end{equation}

\noindent and one coupling parameter, the stiff $K_S$. The transfer matrix method allows for an easy solution. The superspin chain represents the (1+3) Ising model in cases where $J$ locks the plaquette spins into particular configurations, with $p$ either $1$ or $3$.

\par
\begin{figure}
\centering
\includegraphics[width=10cm]{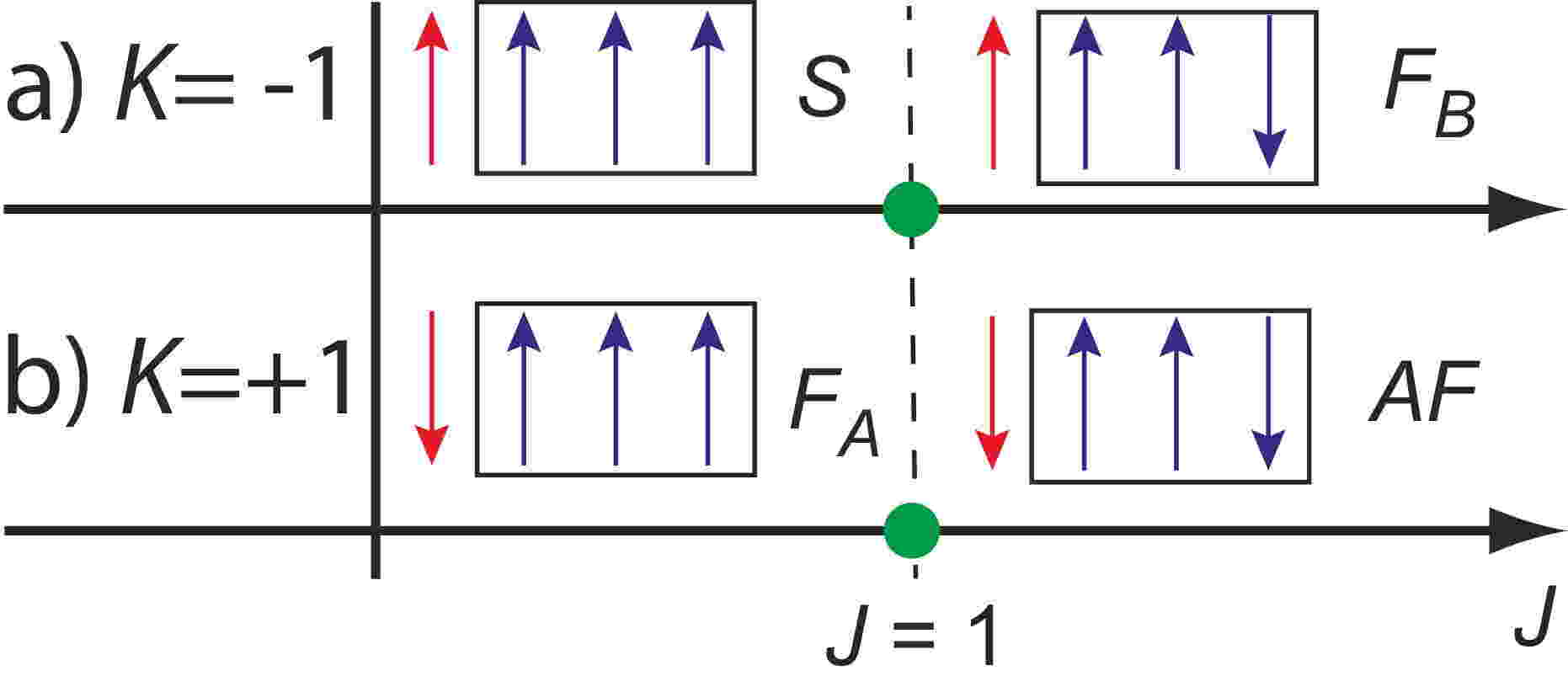}
\caption{Ground state, zero magnetic field, spin configurations as a function of $J$ for a) $K = -1$ and b) $K = +1$. These spin configurations repeat over the entire lattice. Our spin diagrams feature up spins, but since $H = 0$ the configurations with reversed spins are equally probable. $J = 1$ marks the phase boundary for $K = \pm 1$.}
\label{fig:2}
\end{figure}

Let us restrict attention in this paper to zero magnetic field $H = 0$. We consider only the values $K = -1, 0, 1$, which cover the full model [for general $K$, the mapping $(\beta\rightarrow\beta /|K|, H\rightarrow H |K|, J\rightarrow J |K|)$ leaves $\phi$ invariant]. The ground state spin configurations for $K = \pm 1$ are shown in Figure \ref{fig:2}. There is a saturated ferromagnetic state $S$, with all of the spins up, a ferrimagnetic state $F_A$, with all three plaquette spins up, and the single Ising spin down, a ferrimagnetic state $F_B$, with frustrated plaquette spins (two up and one down, with the down spin in any of the three positions), and the single Ising spin directed with the plaquette majority spins, and an antiferromagnetic state $AF$, with frustrated plaquette spins, and the single Ising spin directed with the plaquette minority spin. Appropriate wavefunction symmetrization was done when combining the three spins in every plaquette \cite{Antonosyan2009}. Details involve paired $F_B$ and $AF$ ground states. These show up in the transfer matrix elements Eqs. (\ref{20})-(\ref{40}), but do not figure into the present discussion. The $F_B$ and $AF$ phases have zero magnetic field $s = \mbox{ln}\,3$ as $T\rightarrow 0$, due to frustration.

\par
If $|J/K|$ is large, then the spins in each plaquette lock into place with each other, according to the sign of $J$, as in Fig. \ref{fig:2}. One expects the (1+3) Ising chain to conform to the superspin chain, according to the sign of $J$, with positive $J$ corresponding to the frustrated $p = 1,$ and negative $J$ corresponding to $p = 3$. Otherwise, only the value of $K = K_S$ is important, with variations in $J$ causing little effect. For $K = 0$ we expect paramagnetic behavior, with only small organized fluctuating structure size.

\par
The invariant $R$ results directly from an information theoretic thermodynamic metric, with metric elements $g_{\alpha\beta} = \phi_{,\alpha\beta}$. The coordinates are $(x^1,x^2) = (\beta,h)$, and the comma notation denotes differentiation \cite{Ruppeiner1995, Landau1977}. For the ideal gas, $R = 0$, and near critical points of fluid and spin systems (including critical points at $T = 0$!),

\begin{equation} \xi^d = -\frac{1}{2}R, \label{100}\end{equation}

\noindent where $d$ is the spatial dimensionality (here, $d = 1$) \cite{Ruppeiner1995, Johnston2003}.

\par
Generally \cite{Ruppeiner1995},

\begin{equation} R = \frac{1}{2}\left| \begin{array}{ccc} \phi_{,11}& \phi_{,12}& \phi_{,22}\\ \phi_{,111}&\phi_{,112}&\phi_{,122}\\ \phi_{,112}&\phi_{,122}&\phi_{,222} \end{array}\right| \displaystyle /\left| \begin{array}{cc} \phi_{,11} & \phi_{,12}\\\phi_{,12}&\phi_{,22}\end{array}\right|^2.\label{110}\end{equation}

\noindent $R$ is in units of lattice constants, and depends on derivatives of $\phi$ up to third-order. For fluid systems $R$ was found to be negative when attractive intermolecular interactions dominate, such as near critical points, and positive in cases where repulsive interactions dominate, such as in solids \cite{Ruppeiner2010,Ruppeiner2012a,May2013}. The sign of $R$ has been less explored in spin systems, though recently it was shown that the kagome Ising model (2D) in a magnetic field has $R$ diverging to $\pm\infty$ on opposite sides of the phase transition line ($R<0$ on the ferromagnetic side, and $R>0$ on the antiferromagnetic side) \cite{Mirza2013}.

\par
\begin{figure}
\begin{minipage}[b]{1.5\linewidth}
\includegraphics[width=1.80 in]{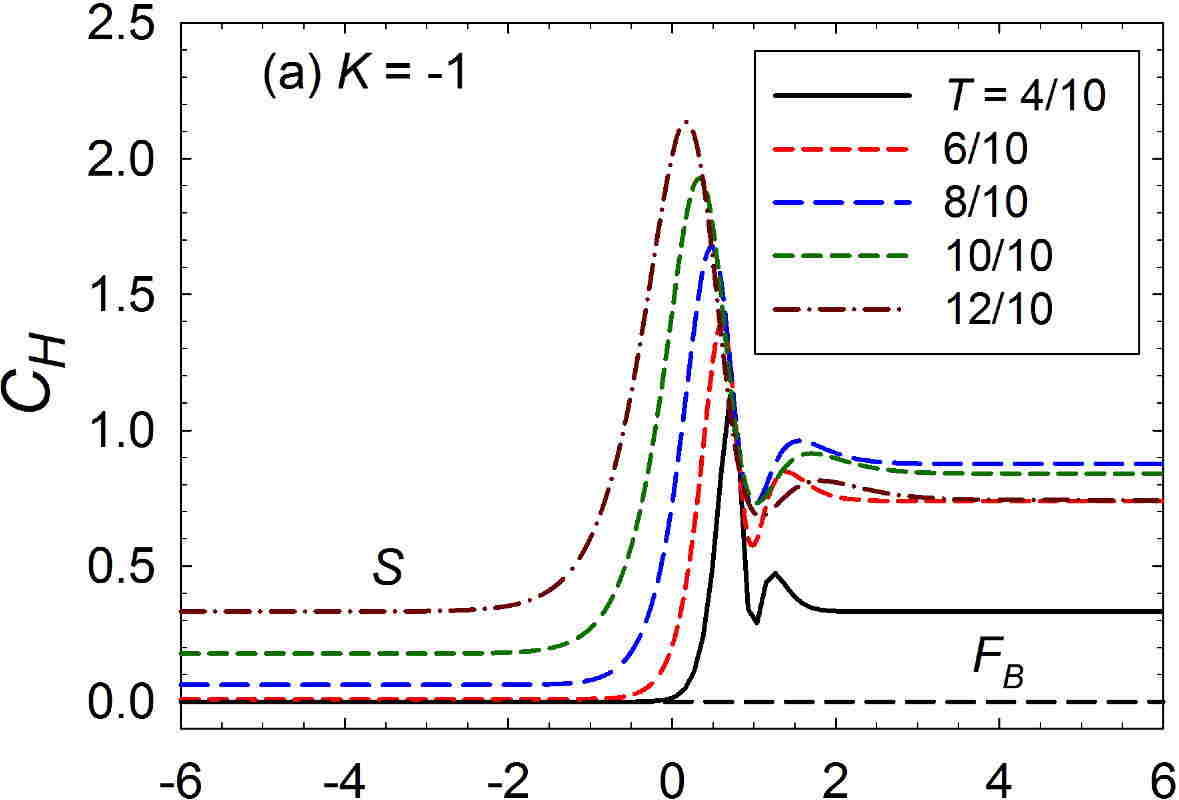}
\hspace{-0.2 cm}
\includegraphics[width=1.80 in]{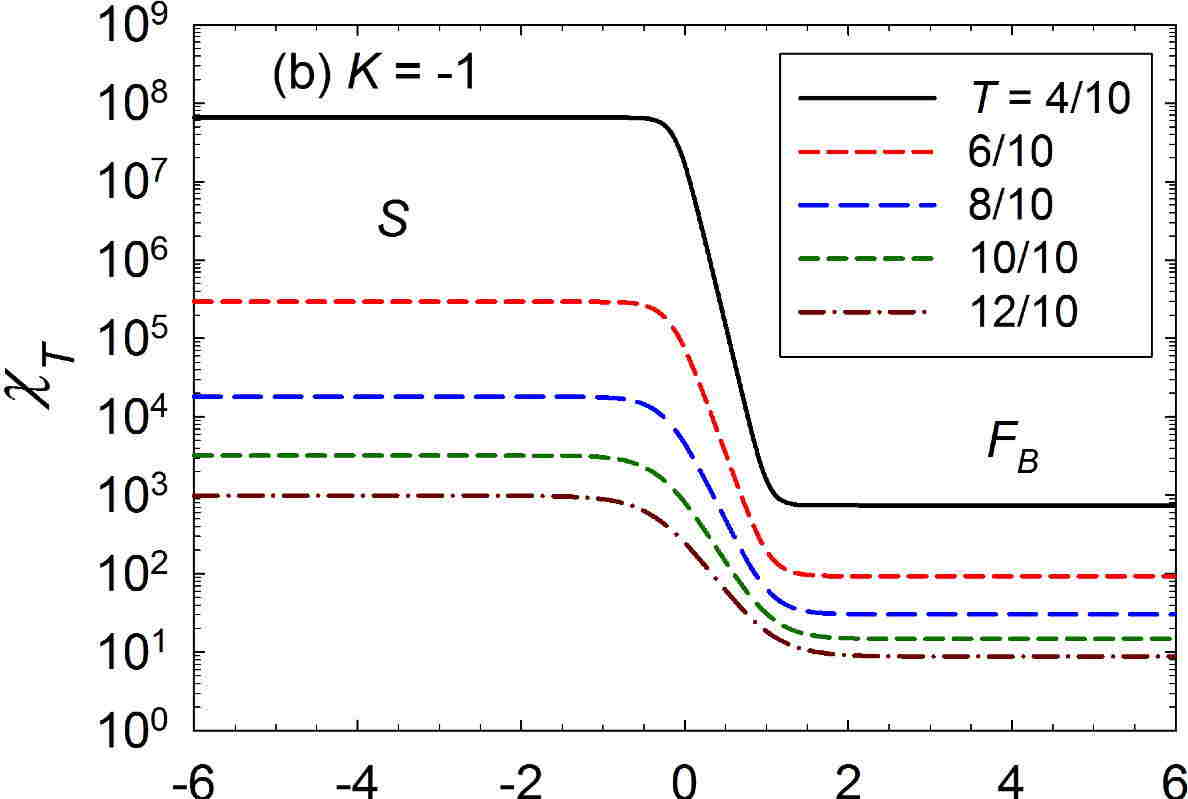}
\hspace{-0.2 cm}
\includegraphics[width=1.80 in]{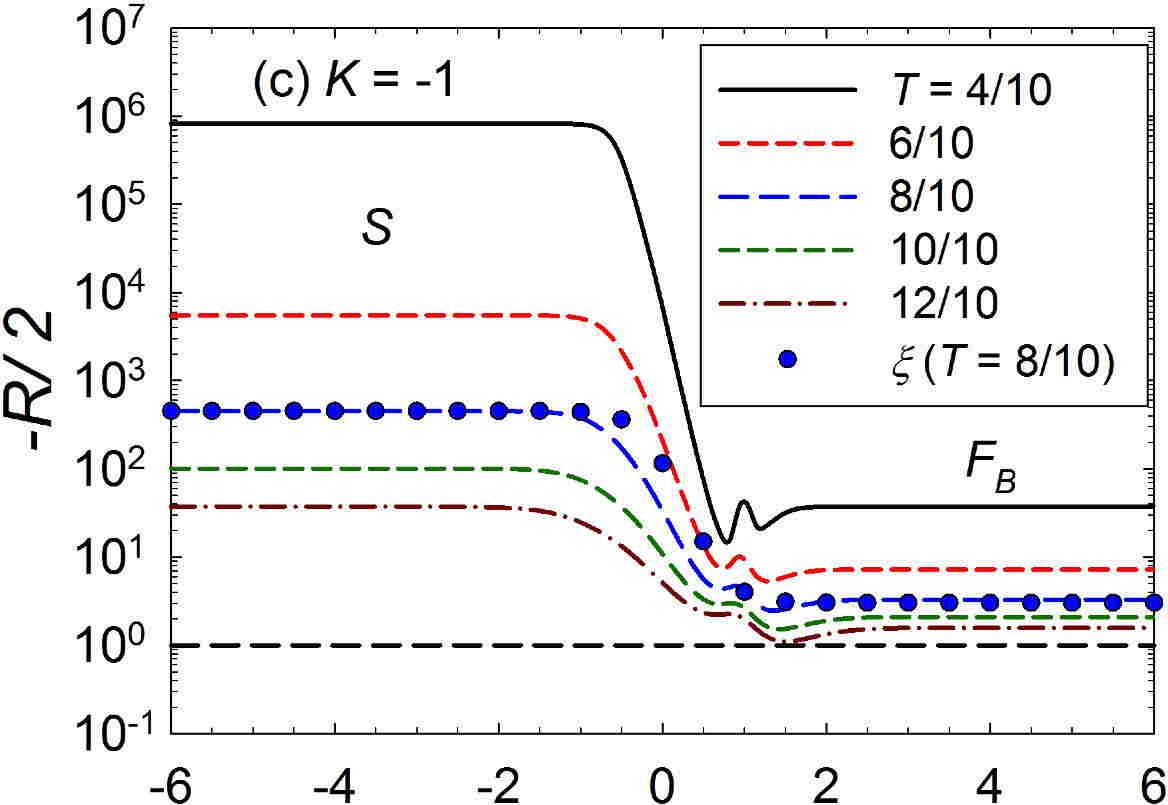}
\vspace{0.1 cm}
\end{minipage}
\begin{minipage}[b]{1.5\linewidth}
\includegraphics[width=1.80 in]{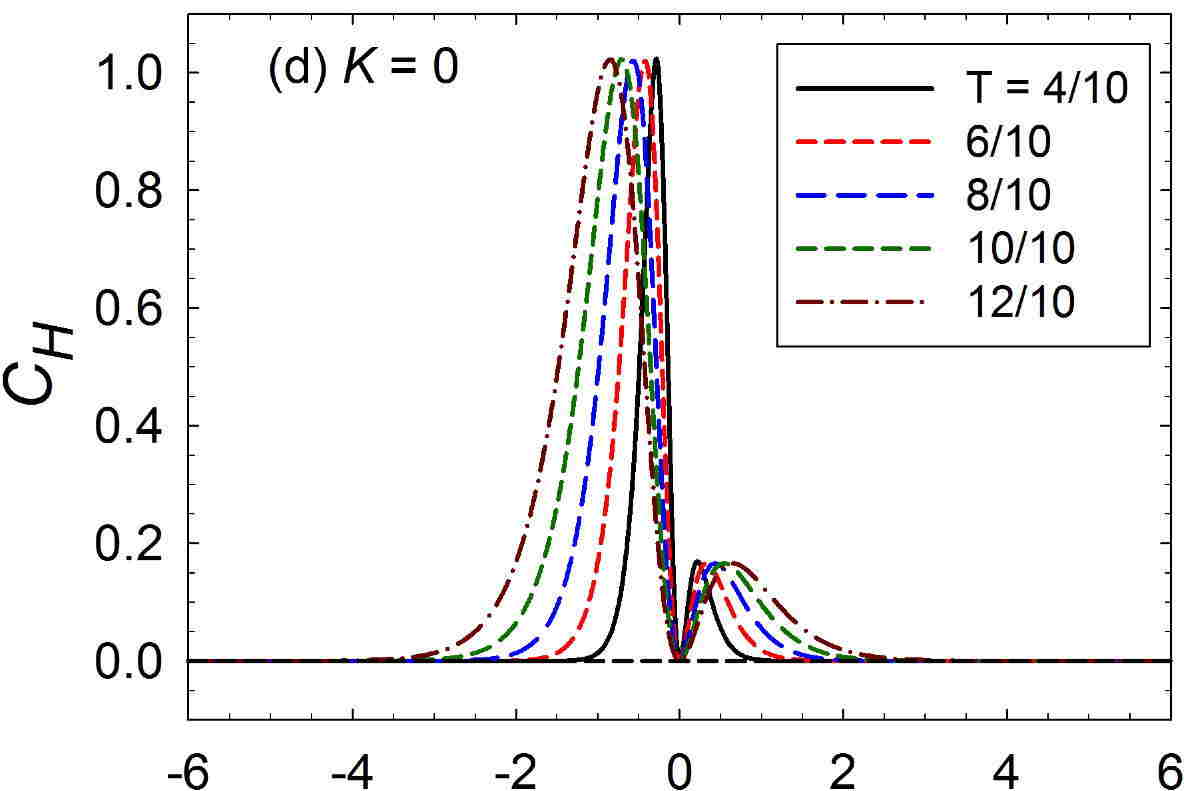}
\hspace{-0.2 cm}
\includegraphics[width=1.80 in]{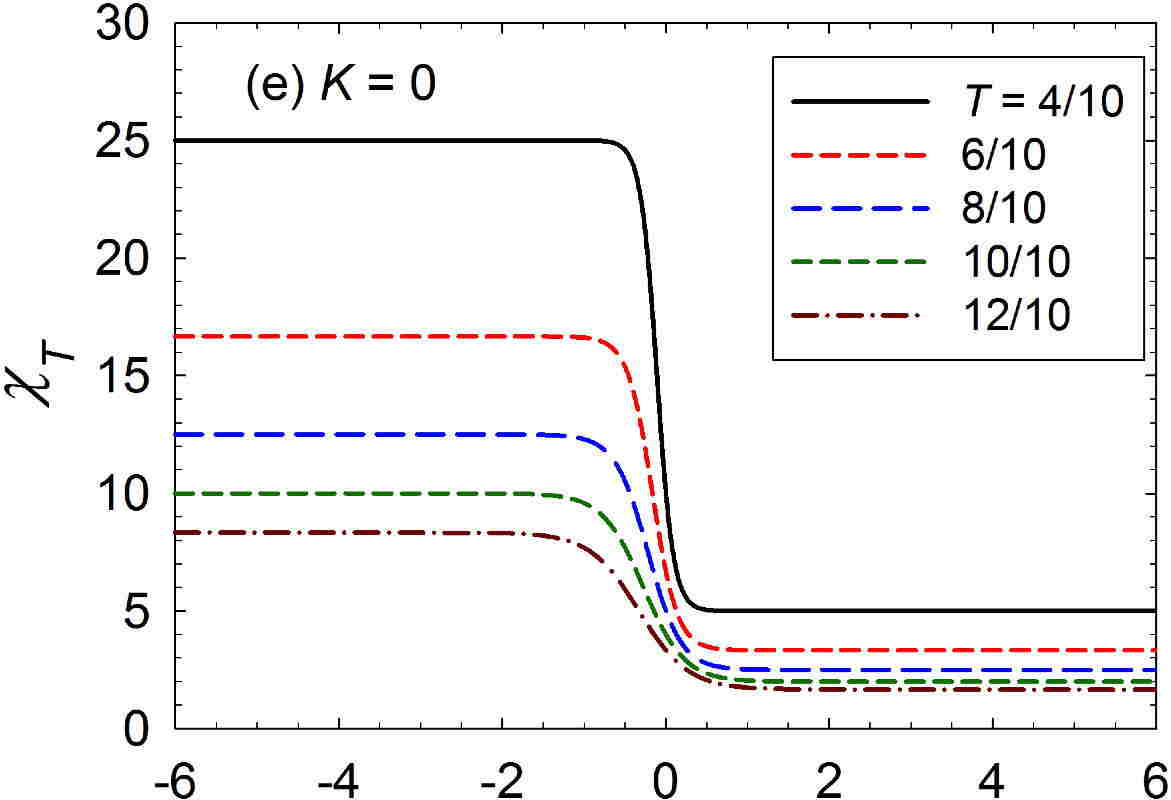}
\hspace{-0.2 cm}
\includegraphics[width=1.80 in]{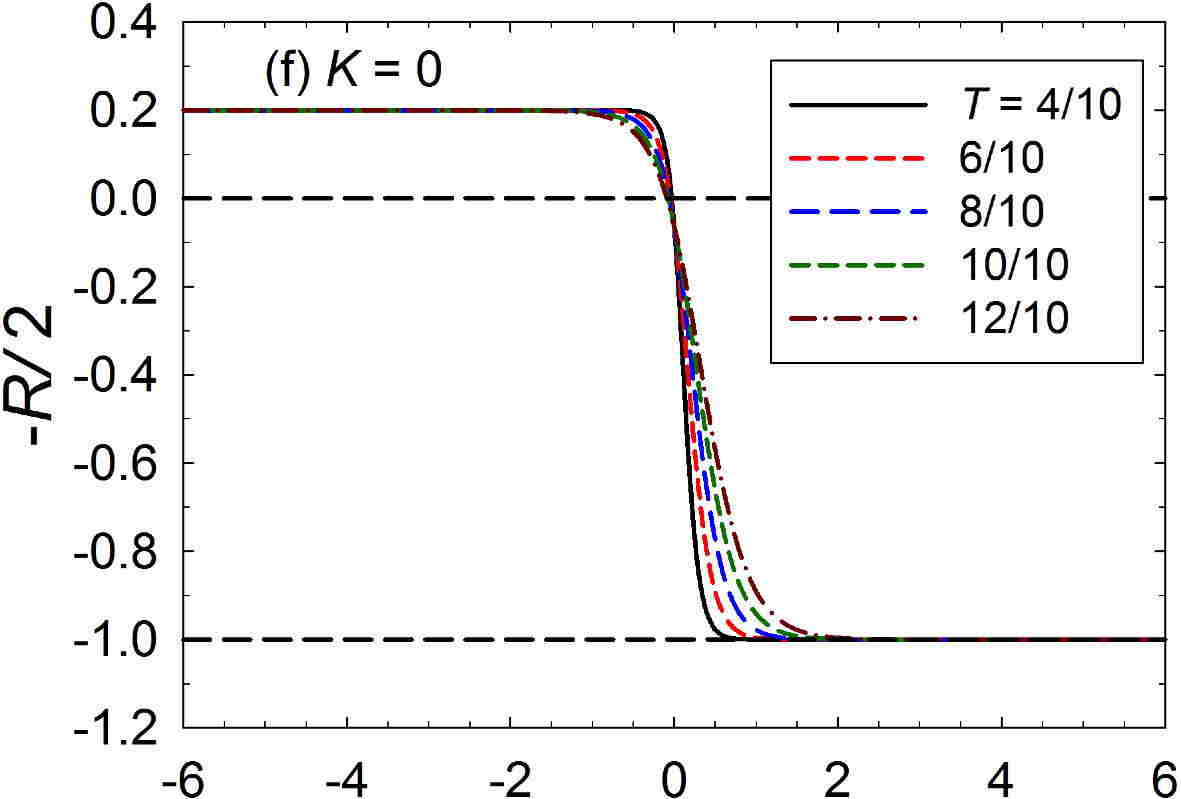}
\vspace{0.1 cm}
\end{minipage}
\begin{minipage}[b]{1.5\linewidth}
\includegraphics[width=1.80 in]{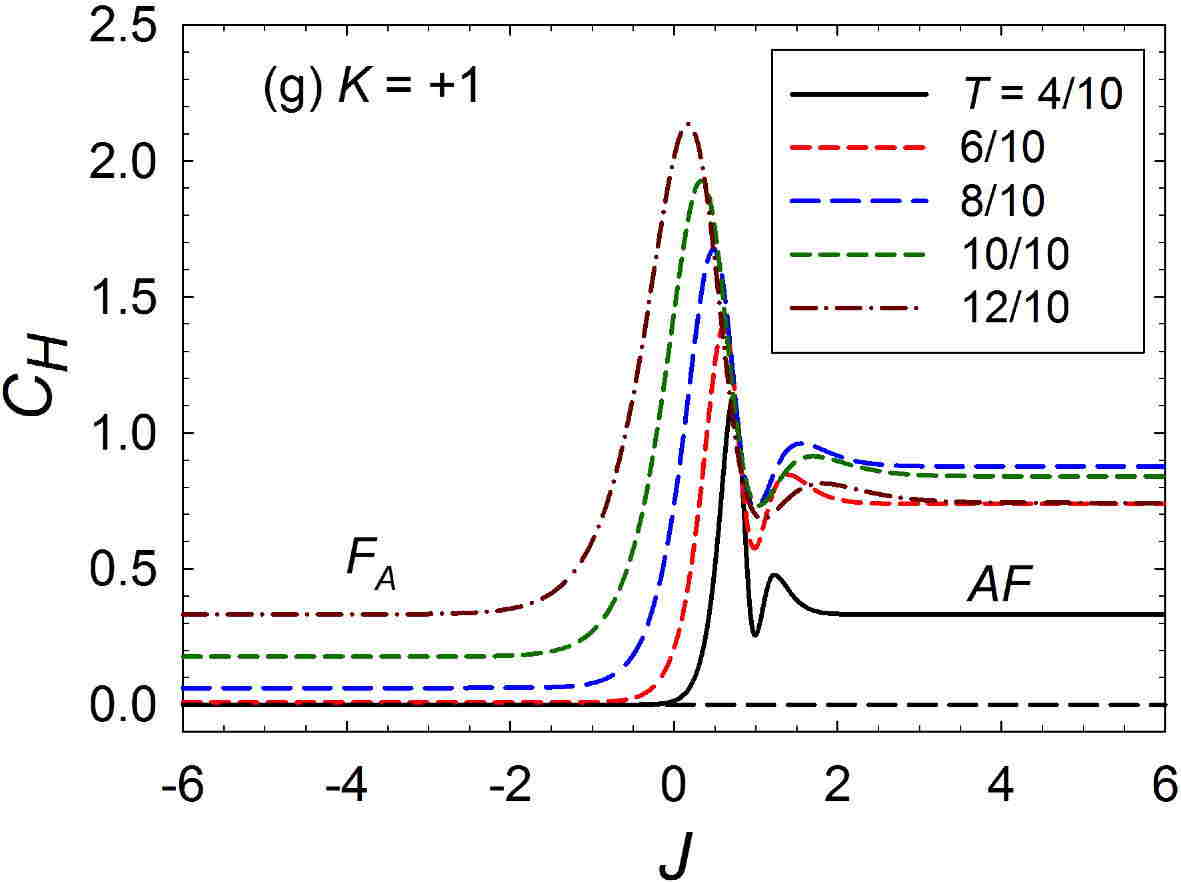}
\hspace{-0.2 cm}
\includegraphics[width=1.80 in]{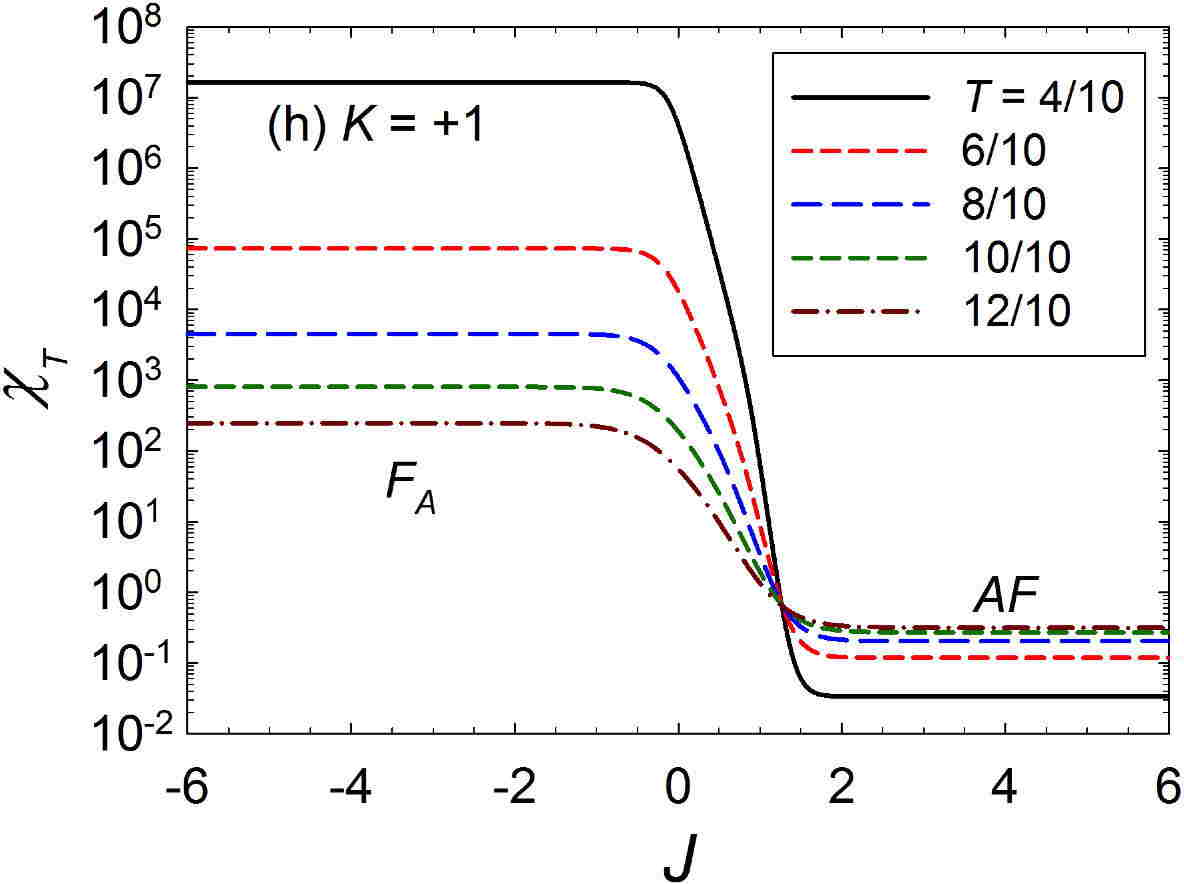}
\hspace{-0.2 cm}
\includegraphics[width=1.80 in]{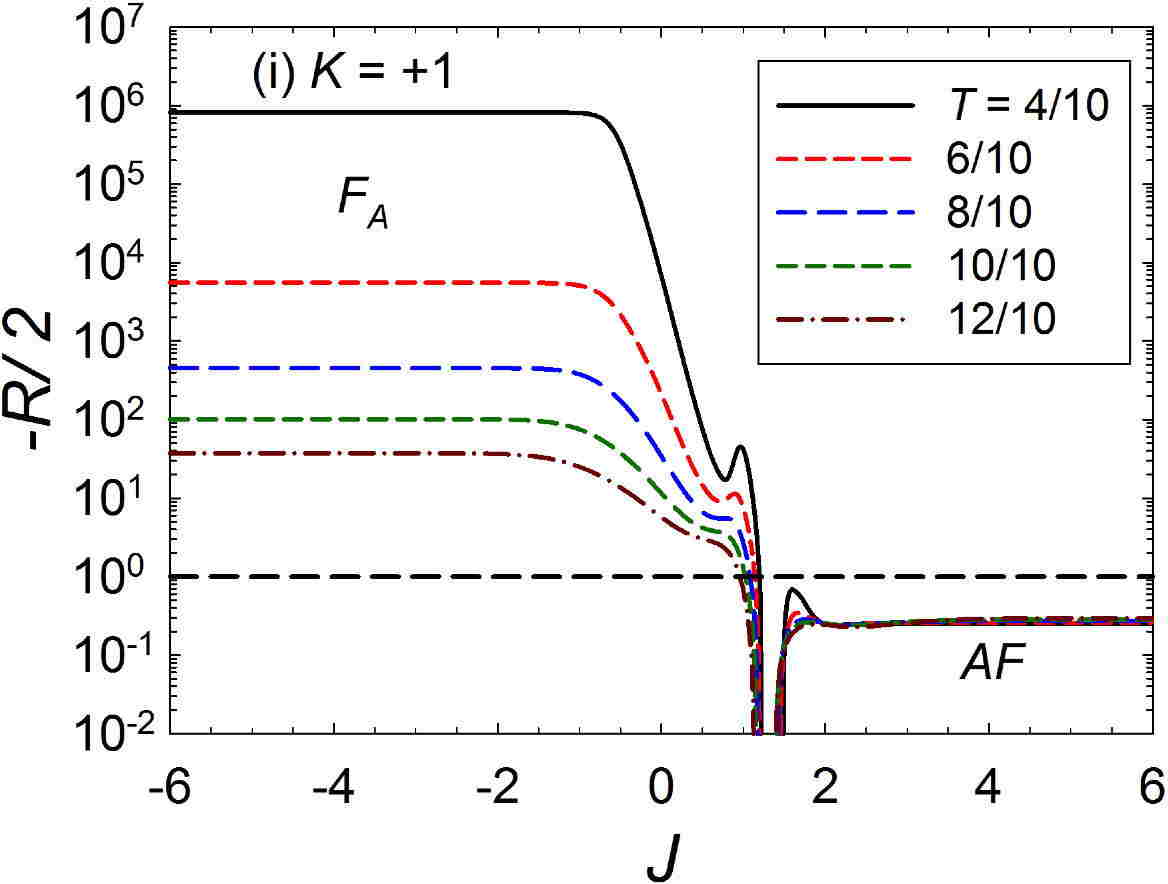}
\end{minipage}
\caption{The heat capacity $C_H$, the magnetic susceptibility $\chi_T$, and the thermodynamic curvature $\xi_R = -R/2$ as functions of $J$ for several values of $T$, and for the three distinct cases $K = -1, 0, +1$. Cases in the plateau regimes, with $J$ not near $1$, are commensurate with the appropriate superspin chains. The dots in (c) show $\xi$ for $T = 8/10$, in good agreement outside the transition regime with the corresponding $\xi_R$. There are negative values of $\xi_R$ in Fig. \ref{fig:3}(i) near $J = 1$, which are omitted on the log scale.}
\label{fig:3}
\end{figure}

Let us define the heat capacity per lattice constant at constant $H$, $C_H = T(\partial s/\partial T)_H$, with entropy per lattice constant $s = \phi-\beta \phi_{,\beta}-h \phi_{,h}$. Also define the magnetic susceptibility $\chi_T = (\partial m/\partial H)_T$, with magnetization per lattice constant $m = -\phi_{,h}$. Figure \ref{fig:3} shows $C_H$, $\chi_T$, and $\xi_R = -R/2$ as functions of $J$ for several values of $T$, and for $K = -1,0,1$. In all cases with large $|J/K|$, these three functions reach asymptotic values independent of $J$, and equal to the corresponding values of the superspin chains with $K_S = K$.

\par
As $J$ increases from very negative values, and reaches the neighborhood of $J = 1$ ($J = 0$ for $K = 0$), all three thermodynamic functions go through transitional values as the corresponding superspin value $p$ goes from $3$ to $1$. For $K = \pm 1$, the transition is between the phases shown in Fig. \ref{fig:2}. In the transitional regime, the sloppy parameter $J$ is clearly very relevant to the thermodynamic behavior, and we can expect no concordance with the superspin chain.

\par
For the paramagnetic state $K = 0$, we have $|\xi_R|\le 1$ lattice constants in all cases, as shown in Fig. \ref{fig:3}(f). Such small values for $|\xi_R|$ are characteristic of situations with weak interactions among constituents. For $\{K,J,H\} = \{0,0,0\}$, $\xi_R = -1/16$ for all $T$, leading to the common crossing point shown in Fig. \ref{fig:3}(f). For $(K, J, H)$ all zero, the spins are randomly directed for all $T$, with $s = 4\,\mbox{ln}\,2$, and $C_H = 0$, as shown in Fig. \ref{fig:3}(d). $\chi_T$ shows a contrast between different $K$'s, having diminished values for the paramagnet. Nevertheless, $\chi_T$ diverges ($\propto\beta$) for the paramagnet in the limit $\beta\rightarrow\infty$, in contrast to $|\xi_R|$ which continues to signal that nothing is going on at long lattice distances. For $K = 0$, the nonthermodynamic $\xi = 0$ for all $J$, so clearly the strictly local $J$ by itself does not produce fluctuations with large spatial extent.

\par
For $K = \pm 1$, and for $|J|$ not too small, Fig. \ref{fig:3} shows strong divergences for $\chi_T$ and $\xi_R$ as $\beta\rightarrow\infty$ in the $S$ and $F_A$ states. Weaker divergences are present in the $F_B$ state. $C_H$ is the same for $K = \pm 1$, since both cases have the same entropy function $s = s(T)$. In the transition regime, to the right of the peaks in Figs. \ref{fig:3}(a) and \ref{fig:3}(g), $C_H$ shows a region of nearly temperature independent behavior. For decreasing $J$, values of $\xi_G$ become the same for $K = \pm 1$, as seen in Figs. \ref{fig:3}(c) and \ref{fig:3}(i), reflecting a zero magnetic field symmetry for the $S$ and $F_A$ states. However, this symmetry is not displayed by $\chi_T$.

\par
For $K = +1$, $\chi_T$ in Fig. \ref{fig:3}(h) has the curves crossing near $J = 5/4$, with the crossing depending weakly on $\beta$. $\xi_G$ in Fig. \ref{fig:3}(i) shows negative minima in the transition region on going from the $F_A$ to the $AF$ state. These minima grow deeper as the temperature decreases. Similar behavior was seen in the Takahashi gas, a one-dimensional system of hard rods with both attractive and repulsive interactions, during a pseudo-phase transition from gas-like to liquid-like \cite{Ruppeiner1990}. By the lattice gas analogy (discussed below), the correspondence between these negative $\xi_R$ features is not unexpected. There is no corresponding feature in the transition from the $S$ to the $F_B$ state in Fig. \ref{fig:3}(c).

\par
The best way to characterize divergences as $\beta\rightarrow \infty$ consists of low temperature, zero magnetic field, series expansions in powers of the small parameter $w = e^{-2 p |K|\beta}$. In the $S$, $F_A$, and $F_B$ phases we find that, to leading order, $\xi_R = w^{-1}/4$, with the same divergence for $\xi$, in accord with Eq. (\ref{100}). These series results (independent of $J$) are strong, holding (with $K\pm 1$) for all integer values of $J$ except $J = 0,1$ in the transition region. The corresponding superspin chains have the same series. The absence of $J$ in both $w$ and the series coefficient $1/4$ further illustrate $J$'s irrelevance out of the transition region. To leading order, $\chi_T = 2^p\beta w^{-1}$ for $K = -1$, and $\chi_T = 2\beta w^{-1}$ for $K = +1$, except for $J = 0,1$. These series for $\chi_T$ are not as clean as those for $\xi_R$, but they make the same point about $J$.

\par
\begin{figure}
\begin{minipage}[b]{0.5\linewidth}
\includegraphics[width=2.7in]{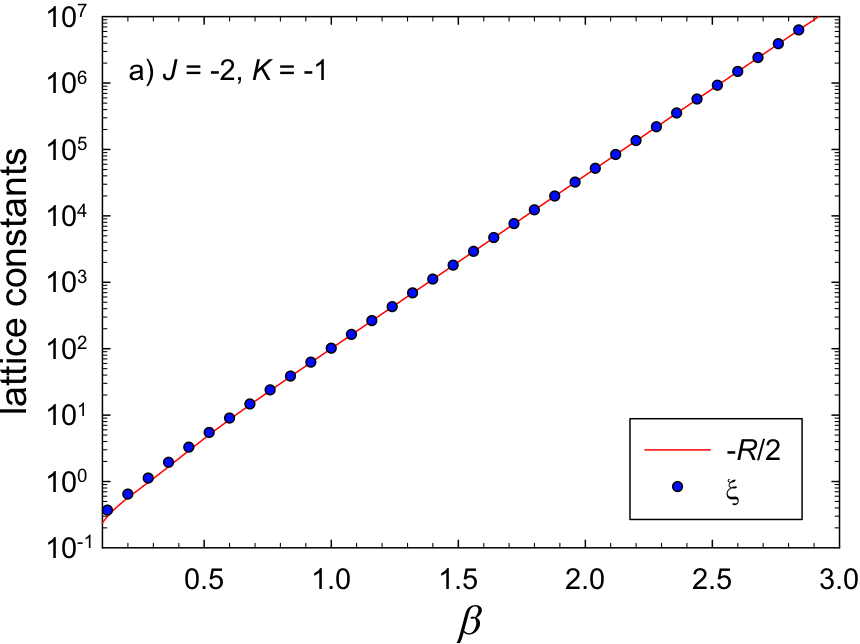}
\end{minipage}
\hspace{0.0 cm}
\begin{minipage}[b]{0.5\linewidth}
\includegraphics[width=2.7in]{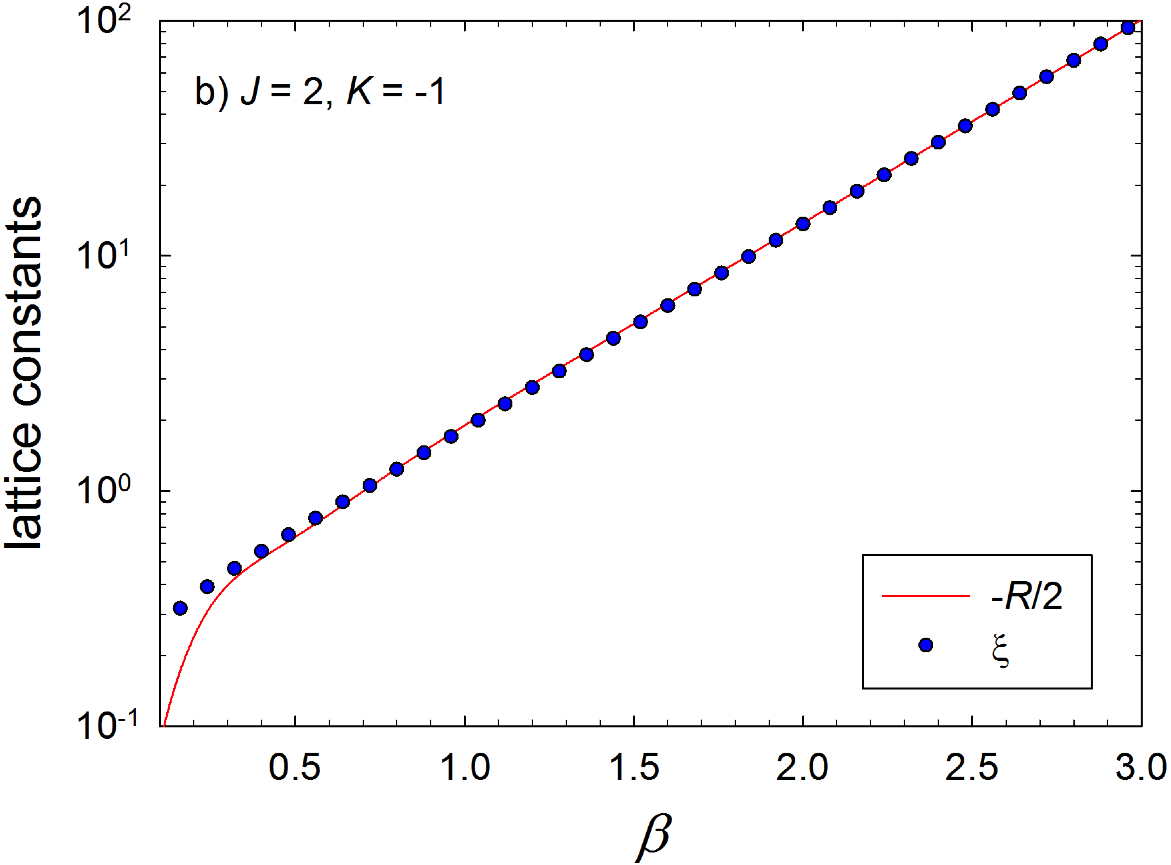}
\end{minipage}
\caption{a) $\xi_R=-R/2$ and $\xi$ for the saturated phase $S$ at zero magnetic field, with $\{J,K\} = \{-2,-1\}$. The agreement between $\xi_R$ and $\xi$ is excellent even to regimes with $\xi_R$ less than a lattice constant. b) the corresponding quantities for the ferrimagnetic phase $F_B$ with $\{J,K\} = \{+2,-1\}$. The agreement between $\xi_R$ and $\xi$ is likewise excellent, except when $\xi_R$ has value a fraction of a lattice site.}
\label{fig:4}
\end{figure}

Supplement the series results for $\xi_R$ and $\xi$ with two examples spanning a range of $\beta$. Figure \ref{fig:4} shows excellent agreement between $\xi_R$ and $\xi$ in both the $S$ and $F_B$ phases, down to length scales less than a lattice constant. The concordance with the corresponding superspin chain (not shown here) is likewise excellent. Outside the transition regime for $J$, the quality of these results is representative of that for other values of $J$, and clearly extends well beyond the critical region.

\par
Let us turn now to the antiferromagnetic $AF$ state. Series expansions show that to leading order in $w$, $\xi_R = 1/4$, and $\chi_T\propto \beta w$, for $J\ge 2$ with $K=+1$, findings evident in Figs. \ref{fig:3}(h) and \ref{fig:3}(i), and in concordance with the corresponding superspin chains. To leading order, $\xi = w^{-1}/4$, also in concordance with the corresponding superspin chain. Clearly, $\xi_R$ is quite different from $\xi$ for antiferromagnets, as $\xi$ diverges in the same way as the ferromagnet, while $|\xi_R|$ has small value. This has long been known for the simple Ising chain \cite{Ruppeiner1981}.

\par
Physically understanding $\xi_R$ for the antiferromagnet benefits from a comparison with fluid systems. Ferromagnetic Ising spin models prefer to have aligned adjacent spins, and critical point properties analogous to those for fluid models. The lattice gas model offers a formal correspondence \cite{Thompson1972}. In the lattice gas model, spin up corresponds to a cell occupied by an atom, and spin down corresponds to an empty cell. Thus, the Ising ferromagnet corresponds to a fluid model with a preference for adjacent occupied cells. Near the critical point, a bunching of atoms, of characteristic size $\xi$, is brought about by the attractive interatomic interactions. The critical point models are characterized by uniformly negative $R$ \cite{Ruppeiner2010}, and by the asymptotic equality Eq. (\ref{100}). The $S$, $F_A$, and $F_B$ states, where all or the majority of spins point in the same direction, and where there is a critical point at $T = 0$, corresponds to a fluid near its critical point. The behavior displayed here is certainly consistent with this expectation. We thus think of ferromagnetic spin interactions as ``attractive.''

\par
We might logically think of the antiferromagnetic interactions as ``repulsive'', with positive $R$, but such thinking is in need of some refinement. Antiferromagnetism tends to have disaligned adjacent spins, corresponding to nearest neighbor atoms avoiding each other in the lattice gas. Outside the transition region for $J$, calculation shows that $|R|$ in the $AF$ phase tends to be uniformly small, of the order of a lattice constant. Although the sign of $R$ for the antiferromagnet is generally negative here, there are cases for this model with the parameter $\Delta\ne 0$ where either sign occurs, though with $|R|$ always of the order of a lattice constant. As was shown by May {\it et al.} \cite{May2013}, solid models tend to have small positive $R$, and condensed liquid states tend to have small $|R|$, with $R$ positive or negative depending on the density. By this measure, the antiferromagnetism here corresponds to the condensed liquid state. In any case, the results we have obtained here for the antiferromagnetic states are fully in accord with expectations from the fluid or solid context.

\par
In conclusion, we have shown that in the zero magnetic field $(1+3)$ Ising chain here, the macroscopic order is connected with the ``stiff'' parameter $K$, whose repeated application connects all of the spins in the chain. The ``sloppy'' parameter $J$, operating only within local spin groups, affects the long-range behavior mostly through its conditioning of the local spin plaquettes for the interaction with $K$. Our analysis emphasized the role of the thermodynamic curvature $R$ at characterizing the resulting magnetism. The ferromagnetic and the ferrimagnetic phases take on negative curvatures, diverging as the correlation length $\xi$ as temperature $T\rightarrow 0$. The antiferromagnet may have positive or negative $R$, with $|R|$ of the order of a lattice constant. We suggest that at zero magnetic field such characteristics, which link directly to fluids or solids through the lattice gas analogy, may be general in spin models. Future research adds a magnetic field ($H\ne 0$), and a full Heisenberg interaction between the plaquette spins ($\Delta\ne 0$). Also most interesting to work out would be a model where the effect of local spin interactions actually average out at the macroscopic level. This would relate our ideas of connecting $R$ from the thermodynamic fluctuating FIM fully to those of Sethna, {\it et al.} \cite{Waterfall2006, Machta2013}.

\par
We thank Vadim Ohanyan for sharing his insight about decorated Ising chains. GR thanks George Skestos for research and travel support, and INFN in Frascati, Italy, where this work was written, for their hospitality.

 \end{document}